\documentstyle[12pt]{article}
\begin{document}
\addtolength{\baselineskip}{.35mm}
\input epsf
\newcommand{\ttau}{r}
\newcommand{\vev}[1]{\langle #1 \rangle}
\def\mapright#1{\!\!\!\smash{
\mathop{\longrightarrow}\limits^{#1}}\!\!\!}
\newcommand{\bigoint}{\displaystyle \oint}
\newlength{\extraspace}
\setlength{\extraspace}{2mm}
\newlength{\extraspaces}
\setlength{\extraspaces}{2.5mm}
\newcounter{dummy}
\newcommand{\be}{\begin{equation}
\addtolength{\abovedisplayskip}{\extraspaces}
\addtolength{\belowdisplayskip}{\extraspaces}
\addtolength{\abovedisplayshortskip}{\extraspace}
\addtolength{\belowdisplayshortskip}{\extraspace}}
\newcommand{\ee}{\end{equation}}
\newcommand{\figuur}[3]{
%\addtocounter{fignum}{1}
%\addcontentsline{lof}{figure}{\protect
%\numberline{\arabic{section}.\arabic{fignum}}{#3}}
%\hspace{-3mm}{\it fig.}\ %\figuurnum.
\begin{figure}[t]\begin{center}
\leavevmode\hbox{\epsfxsize=#2 \epsffile{#1.eps}}\\[3mm]
\bigskip
\parbox{15.5cm}{\small %\figuurnum:
 #3}
\end{center} \end{figure}\hspace{-1.5mm}}
\newcommand{\figuurb}[3]{
%\addtocounter{fignum}{1}
%\addcontentsline{lof}{figure}{\protect
%\numberline{\arabic{section}.\arabic{fignum}}{#3}}
%\hspace{-3mm}{\it fig.}\ %\figuurnum.
\begin{figure}[t]\begin{center}
\leavevmode\hbox{\epsfxsize=#2 \epsffile{#1.eps}}\\[-3mm]
\bigskip
\parbox{15.5cm}{\small #3}
\end{center} \end{figure}\hspace{-1.5mm}}

\newcommand{\fig}{{\it fig.}\ }
\newcommand{\newsection}[1]{
\vspace{15mm}
\pagebreak[3]
\addtocounter{section}{1}
\setcounter{subsection}{0}
\setcounter{footnote}{0}
\noindent
{\Large\bf \thesection. #1}
\nopagebreak
\medskip
\nopagebreak}
\newcommand{\newsubsection}[1]{
\vspace{1cm}
\pagebreak[3]
\addtocounter{subsection}{1}
\addcontentsline{toc}{subsection}{\protect
\numberline{\arabic{section}.\arabic{subsection}}{#1}}
\noindent{\bf %\thesection.
\thesubsection. #1}
\nopagebreak
\vspace{2mm}
\nopagebreak}
\newcommand{\ba}{\begin{eqnarray}
\addtolength{\abovedisplayskip}{\extraspaces}
\addtolength{\belowdisplayskip}{\extraspaces}
\addtolength{\abovedisplayshortskip}{\extraspace}
\addtolength{\belowdisplayshortskip}{\extraspace}}
\newcommand{\one}{{\bf 1}}
\newcommand{\zbar}{\overline{z}}
\newcommand{\ea}{\end{eqnarray}}
\newcommand{\is}{& \!\! = \!\! &}
\newcommand{\hf}{{1\over 2}}
\newcommand{\del}{\partial}
%
% 2 by 2 matrices
%
\newcommand{\twomatrix}[4]{{\left(\begin{array}{cc}#1 & #2\\
#3 & #4 \end{array}\right)}}
\newcommand{\twomatrixd}[4]{{\left(\begin{array}{cc}
\displaystyle #1 & \displaystyle #2\\[2mm]
\displaystyle  #3  & \displaystyle #4 \end{array}\right)}}
\newcommand{\low}{{{\rm {}_{IR}}}}
\newcommand{\hi}{{{\rm {}_{UV}}}}
\newcommand{\hilo}{{{}_{{}^{\rm {}_{UV/IR}}}}}
\newcommand{\ie}{{\it i.e.\ }}
\newcommand{\gbar}{{\overline{g}}}
\newcommand{\half}{{\textstyle{1\over 2}}}
\newcommand{\tfrac}{\frac}
\newcommand{\XX}{\dot{g}}
\newcommand{\XXX}{{\mbox{\small \sc X}}}
\newcommand{\xx}{p}
\newcommand{\kappaf}{\kappa_{{}_{\! 5}}}
\newcommand{\CCC}{{\mbox{\large $\gamma$}}}
\renewcommand{\thesubsection}{\arabic{subsection}}
\renewcommand{\footnotesize}{\small}
\begin{titlepage}
\begin{center}

{\hbox to\hsize{
\hfill PUPT-1925}}

\bigskip

\vspace{6\baselineskip}

{\large \bf Supersymmetry at Large Distance Scales}
\bigskip
\bigskip
\bigskip

{Herman Verlinde}\\[1cm]

{ \it Physics Department, Princeton University, Princeton, NJ 08544}\\[5mm]

\vspace*{1.5cm}

{\bf Abstract}\\

\end{center}
\noindent
We propose that the UV/IR relation that underlies the AdS/CFT duality may provide
a natural mechanism by which high energy supersymmetry can have large
distance consequences. We motivate this idea via (a string realization
of) the Randall-Sundrum scenario, in which the observable matter is
localized on a matter brane separate from the Planck
brane. As suggested via the holographic interpretation of this
scenario, we argue that the local dynamics of the Planck brane
-- which determines the large scale 4-d geometry -- is
protected by the high energy supersymmetry of the dual 4-d theory. 
With this assumption, we show that the total vacuum energy
naturally cancels in the effective 4-d Einstein equation. This cancellation 
is robust against changes in the low energy dynamics on the matter brane, which
gets stabilized via the holographic RG without any additional fine-tuning.

\end{titlepage}

\newpage
\newcommand{\UU}{S_E}%{{\cal U}}
\newcommand{\cto}{\mu}

\newsubsection{Introduction}

The observed smallness of the cosmological constant requires a
remarkable cancellation of all vacuum energy contributions \cite{weinberg}.
Supersymmetry seems at present the only known symmetry that could
naturally explain this cancellation, but thus far no mechanism for
supersymmetry breaking is known that would not destroy this
property. Nonetheless, in searching for a possible resolution of the
cosmological constant problem, it seems natural to include
supersymmetry as a central ingredient. What would then be needed,
however, is a mechanism -- some UV/IR correspondence -- by which the
short distance cancellations of supersymmetry can somehow be
translated into a long distance stability of the cosmological
evolution equations. In this paper we would like to propose a possible
candidate for such a mechanism. The basic idea is as follows.

\smallskip

Consider a string realization \cite{hv}\cite{cpv} of the Randall-Sundrum
compactification scenario \cite{rs1}. In such a compactification,
space-time consists of a slice of $AdS_5$ (times some small compact
5-manifold) bounded by two brane-like structures, which we will call
the Planck and matter brane, respectively. Both branes must be thought
of as {\it effective} descriptions of more elaborate structures,
defined by the full high energy string theory \cite{cpv} and the low energy
quantum field theory \cite{post}, respectively.  Now, via the
holographic UV/IR duality \cite{uvir}\cite{uvir2} 
of the AdS/CFT correspondence
\cite{adscft}\cite{gkp}\cite{ew}, 
we may identify the 5-d physics at different radial
locations in the bulk region with 4-d physics at corresponding
intermediate energy scales. The fact that the matter degrees of
freedom are localized on a matter brane separate from the Planck
brane, is therefore just a reflection of the fact that they spend most
of their time at much lower energy scales than the Planck scale. This
4-d matter still feels ordinary long distance gravity via its
interaction with the 5-d bulk \cite{rs1}.

\smallskip

The two ingredients that we would like to add to this scenario 
are the following:

\vspace{-3mm}

\begin{itemize}

\item{Since all observable (non-supersymmetric) matter is localized at
the matter brane, it seems an allowed assumption that the physics on
and near the Planck brane is supersymmetric, at least to a very good
approximation.  In particular, the classical action that describes the
effective dynamics of the Planck brane could reasonably be taken to be
of a particular supersymmetric form.  Via the AdS/CFT dictionary, this
assumption can be thought of as the holographic image of requiring
4-d high energy supersymmetry.}

\item{Since almost all of the 5-d volume is contained in the Planck 
region, the large distance structure of the effective 4-d geometry
will be determined by the shape of the Planck brane. Locality of the
5-d supergravity in turn implies that this shape is determined by
{\it local 5-d equations} of motion, satisfied by the 5-d supergravity
fields in the direct neighborhood of the Planck brane.}

\end{itemize}

Both these assumptions, if indeed realizable, are most likely valid
only within a certain level of approximation.  It is clear however
that, if both are truly valid, combined they imply that the equations
of motion that determine the large scale 4-d geometry can indeed be
protected by supersymmetry.  This possible reappearance of
supersymmetry at large distance scales can be seen as related to the
fact that -- unlike in the conventional non-compact set-up -- the
UV/IR mapping of the AdS/CFT correspondence now acts on {\it one
single space-time} that combines both the 4-d boundary field theory
and the 5-d bulk gravity. Via this UV/IR duality, the infra-red bulk
region of the AdS-space near the Planck brane becomes the natural home
base for both the shortest and longest distance physics.

\smallskip

In the following, we will try to test the consistency of these
two assumptions. To this end, we will address the following obvious
and most serious counter-argument. Intuitively, one would expect that
the low energy matter sector on the matter brane will produce some
quite arbitrary effective tension, that (without some unnatural or non-local 
fine-tuning) is expected to induce a non-zero
cosmological constant for the total 4-d effective field theory.  If
indeed present, its backreaction would curve the Planck brane and
consequently break its supersymmetry.

\smallskip

A different version of the same objection is that the AdS/CFT
dictionary tells us that the normal variations of the local
supergravity fields near the Planck brane in fact know about low
energy quantities of the dual field theory, such as vacuum expectation
values, etc.  In particular, the normal variation of the bulk metric
(or more precisely, the {\it extrinsic} curvature at the Planck brane
\cite{bk}) knows about the full vacuum energy produced by the low 
energy field theory.  It would seem quite unnatural to expect that 
the Planck brane dynamics could be chosen such that, without any 
pre-knowledge of the
IR dynamics, it exactly cancels this matter contribution to the vacuum
energy.

\smallskip

In the following sections we will describe a mechanism that will 
neutralize this counter-argument.  In the final section we address some other
aspects of our proposal, and discuss its relation with other recently
proposed scenarios \cite{adks}\cite{kss}.

\figuur{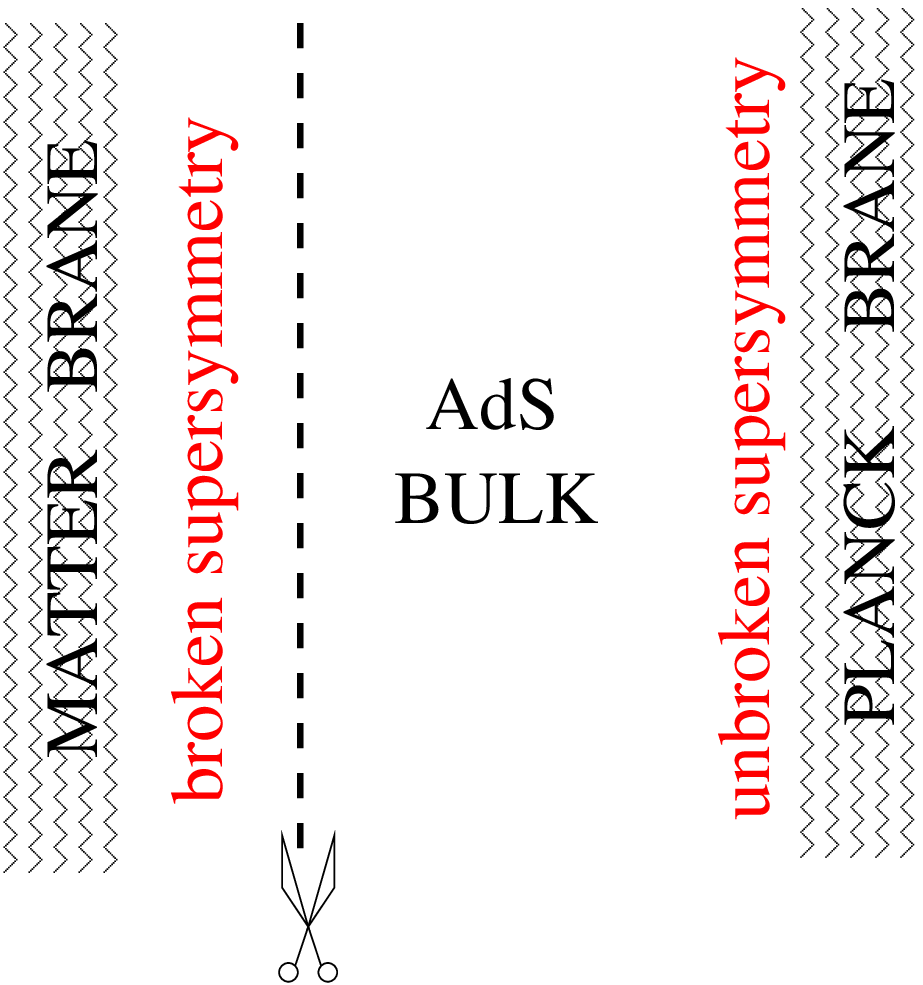}{5.5cm}{{\bf Fig 1.} \ {\it The 5-d geometry is bounded by
two brane-like structures, the Planck and matter brane. 
The physics near the Planck brane is assumed to be supersymmetric.
%Their respective microscopic descriptions are given in the main text.   
The line of separation between the matter brane and the 5-d bulk is 
indicated by the dashed line.\\[-11mm]}}

\vspace{-6mm}

\newsubsection{Set-up}

We first briefly state our assumptions about the physics inside
the three different regions -- the bulk region, the matter brane,
and the Planck brane.\\

\noindent
{\it The bulk supergravity}

The 5-dimensional bulk region is a negatively curved space, with a
varying 5-d cosmological constant $V(\phi)$, typically of planckian
magnitude and dependent on various bulk matter fields
$\phi$. The metric thus takes a warped form, such that the ratio between
the warp factor on the Planck and matter brane matches the hierarchy
between the Planck and typical matter scale \cite{rs1}. The classical equations
of motion for the bulk region are prescribed by a 5-d gauge supergravity
action, of the schematic form (omitting fermionic fields)
\be
\label{bulk}
S_{\rm bulk} = \int \! d^5 x \, \sqrt{-G}\, ( V(\phi)
+   \kappa R  - {1\over 2} (\nabla\phi)^2 ).
\ee
%Here $\phi$ is short-hand for all 5-d matter multiplets. \\

\noindent
{\it The matter brane}

The matter brane hosts all visible matter. 
It typically represents a strongly curved or even
singular region of the 5-d geometry. Because the supergravity
approximation breaks down inside this region, we will introduce 
an artificial line of separation between the
matter brane and the bulk, located at some arbitrary scale
(see fig 1). On this line, we consider the
values of the 5-d supergravity fields $(g,\phi)$. Here $g$ 
is short-hand for the 4-d metric $g_{\mu\nu}$
and $\phi$ determine the couplings and expectation values of the matter 
theory.  The matter brane effective action 
\be
\label{gamma}
\Gamma_{\! \rm matter\, }(g,\phi) 
%\, = \, \int\! d^4x \, \sqrt{-g} \,{\cal L}(g, \phi).  
\ee 
is obtained by integrating out all degrees of freedom to the left of 
the dividing line, with fixed boundary values for $g$ and $\phi$. 
%At low energies, we expect that ${\cal L}(g,\phi)$ can be expanded as 
%\be {\cal L}(g,\phi) = U(\phi) +
%\Phi(\phi) R + G(\phi) (\nabla \phi)^2 + \ldots \ee 
Since supersymmetry is broken in this sector, we will not require any
special symmetries of this effective action.
Moreover, we will in principle allow the
matter brane to have a finite temperature and/or matter density, which
may vary with time and position; $\Gamma$ will then 
contain specific terms whose variation with respect to $g_{\mu\nu}$ will 
equal the corresponding stress-energy contributions.

\smallskip

The effective action $\Gamma_{\rm matter}$ will not be completely
arbitrary, however: consistency of the geometric set-up requires that
the total partition function of the bulk and matter region does not
depend on the location of the artificial dividing line between the two
regions. In AdS/CFT dual terms, this means that we imagine that around
the energy scale corresponding to this line, the matter theory has
become equivalent to the exact holographic dual of the 5-d bulk
supergravity. Shifting the position of the line then corresponds to
changing some arbitrarily chosen RG scale \cite{eh}. We will formulate this 
invariance requirement in more detail in section 5.

\smallskip

\noindent
{\it The Planck brane}

The region near the Planck brane represents the internal compactification 
geometry of the full string theory outside the AdS-like region \cite{hv}.  
In accordance with the holographic interpretation of the 5-d geometry,
its internal structure is directly probed only by Planck scale 4-d
physics.  In our scenario, this physics is assumed to be
supersymmetric. To make this symmetry manifest, we will now postulate, as an
{\t effective} description of the Planck brane brane dynamics, the following
special boundary action
\be
\label{fine}
S_{ \rm planck\, }(g,\phi) \, =\, \int \! d^4 x \sqrt{-{g}} \, W(\phi).
\ee
Here $W(\phi)$ is the superpotential of the 5-d supergravity, 
related to $V(\phi)$ via
\be
\label{finet}
V(\phi) =
{1\over 2} (\partial_\phi W(\phi))^2  -{1\over 3} W(\phi)^2.
\ee
In particular, we will be assuming that  the 
Planck brane will not carry any independent matter density on its own world 
volume.\footnote{This is a simplification, that in effect amounts to
a restriction on the 4-d physics, namely that it is always sub-Planckian. This 
in particular means that we will be ignoring the possible stress-energy contributions 
of gravitationally collapsed matter inside 4-d black holes.}
As we will see shortly, the choice of action (\ref{fine}) implies that the
Planck brane geometry satisfies an effective 4-d Einstein equation with
vanishing cosmological constant.  Ultimately, it will therefore be
important to determine how well protected (\ref{fine})-(\ref{finet}) 
are against quantum corrections induced via the presence
of the non-supersymmetric matter brane. For now, however, we will
simply {\it define} the classical action of our model via
(\ref{fine})-(\ref{finet}). 
%\end{itemize}

\newsubsection{Equation of Motion I}

The boundary conditions that follow from (\ref{fine}) are %\cite{pk}\cite{ssg}
\be
\label{twee}
\theta_{\mu\nu} \! - \theta g_{\mu\nu} = W(\phi) g_{\mu\nu}\qquad \qquad \
\partial_n \phi = \partial_\phi W(\phi),
\ee
where $g_{\mu\nu}$ denotes the induced metric and 
$\theta_{\mu\nu}$ the extrinsic curvature of the Planck brane;
$\partial_n$ the derivative normal to its world-volume. Now
the normal component of the 5-d bulk Einstein equation, when written in 
terms of 4-d geometric quantities, reads
\be
\label{drie}
{1\over 4}(\theta^2 - (\theta_{\mu\nu})^2)
- {1 \over 2}(\partial_n \phi)^2 + 
V(\phi) + \kappa R - {1\over 2}(\nabla\phi)^2 = 0.
\ee
Here $\nabla$ and $R$ are the 4-d gradient and curvature scalar.
When combined with (\ref{twee}) and (\ref{finet}), we deduce from
(\ref{drie}) that on the Planck brane world volume
\be
\label{vier}
\kappa R - {1\over 2} (\nabla\phi)^2 = 0.
\ee
The classical supersymmetric Planck brane geometry thus solves the trace of
the 4-d Einstein equations, with possibly non-zero matter density,
but always with zero cosmological constant. This result holds
independently of what happens inside the bulk: possible corrections 
to (\ref{vier}) can only arise from local physics that happens
at or near the location of the Planck brane.
In the following sections we will present a rederivation of (\ref{vier})
within the quantum context.

\newsubsection{Partition Function}

Consider the total quantum mechanical partition function of
our system.  We can first divide it into three parts, corresponding to the
three sub-regions. The bulk supergravity partition function can best
be thought of as an evolution operator $\widehat{\cal U}$ that
describes the propagation through the bulk from the matter to the
Planck brane. Its matrix element between eigen states $|g,\phi\rangle$
of the boundary fields has the formal path integral expression
\be
\langle g_1,\phi_1|\, \widehat{\cal U} \,  | g_2, \phi_2 \rangle \, = \, 
\int\limits_{(g_1,\phi_1)}^{(g_2,\phi_2)}\!
\![dG][d\phi] \; e^{\textstyle {i\over \hbar} S_{\rm bulk}(G,\phi)}.
\ee
Since in our set-up the matter brane is defined as the region
left to some quite arbitrary line of separation with the bulk (see fig
1), its partition function is indeed most appropriately thought of as
a wave-function
\be
\langle \Psi_{\! \rm matter\, }|g_1,\phi_1 \rangle= \, 
e^{\textstyle \, {i\over \hbar} \, \Gamma_{\! \rm matter\, }(g_1,\phi_1)}. 
\ee
Similarly we can define
\be
\label{psipl}
\langle g_2,\phi_2| \Psi_{\! \rm planck\, }\rangle = \, 
e^{\textstyle \, {i\over \hbar} \, S_{ \rm planck\, }(g_2,\phi_2)}. 
\ee
The total partition function is obtained by taking the inner-product
of these two wave-functions, with the bulk evolution operator inserted
in between
\be
\label{z}
Z\, = \, \langle \Psi_{\! \rm matter} | \, \widehat{\cal U} \,
| \Psi_{\! \rm planck}
\rangle.
\ee
Inserting the above definitions, this matrix element represents the
complete functional integral over all matter and gravity fields. 
We will imagine that it can be given a well-defined definition via the
underlying fundamental string theory.

\newsubsection{Holographic RG}

Now let us consider the evolution of the matter state under local
variations of the position of the dashed line in fig 1. Thinking about
the radial direction as a Euclidean time directon, we may identify the
corresponding Hamilton operator $\hat{\cal H}$ of the bulk
supergravity as the generator of these variations. However, as 
in any gravity theory, physical wave functions should be
invariant under local variations of the particular time-slicing.  We may
write this condition as
\be
\label{cst} 
\hat{\cal H} \; \,{\widehat{\cal U}}\, | \Psi_{\! \rm matter\, }
\rangle \, = \, 0.
\ee
The Hamilton operator associated with the bulk action (\ref{bulk})
reads
\be
\label{fff}
\hat{\cal H} \; = \; 
 {{\hbar^2 \over \sqrt{-g} }}
( \, {1 \over 3} \Bigl(%\hbar
\frac{\delta  \ } 
{\delta g^\lambda_\lambda}\Bigr)^2\!\! -\!
 \Bigl(%\hbar
\frac{\delta \ \ }{\delta g^{\mu\nu}}\Bigr)^2\!
\! -  \! \frac{1}{2} 
\Bigl(%\hbar 
{\delta \  \over \delta \phi}\Bigr)^2 )
+  \sqrt{-g} (  V(\phi) \! +  \kappa R  - \! \textstyle{{1\over 2}} 
(\nabla \phi)^2 ) .
\ee
Classically, the condition $\hat{\cal H} = 0$ is simply the 5-d bulk
Einstein equation (\ref{drie}). Indeed, we emphasize that
eqn (\ref{cst}) is not some special symmetry requirement on the
matter sector, but a property of {\it any} state that has
traveled through the bulk supergravity region. 

\smallskip

After applying the AdS/CFT dictionary, our geometric set-up is equivalent to assuming that the
complete matter theory above some energy scale -- corresponding roughly to the line of
separation between the matter brane and the bulk -- becomes equivalent to the exact holographic
dual of the bulk supergravity.  From this 4-d perspective, the relation (\ref{cst}) acquires a
new meaning as the evolution of the matter partition function under the holographic RG flow
\cite{eh,rgt}.  Note however, that for finite rank $N$ and gauge
coupling, the bulk theory is a complete
string theory with finite string length and string coupling.  The
constraint (\ref{cst}) for finite $\hbar$ includes $1/N$ effects, but the Hamiltonian
(\ref{fff}) will receive various string corrections.  We expect however that these will not
significantly alter the basic form of the RG equation \cite{kv}.

\newsubsection{Equation of Motion II}

In our set-up, the boundary state 
at the Planck brane has been chosen to be invariant under 4-d
global supersymmetry transformations,
$Q_\alpha  |\Psi_{\rm planck}\rangle = 0$.
Using the explicit form (\ref{psipl}) of  
$|\Psi_{\rm planck}\rangle$ and the
classical relation (\ref{finet}), we 
find that
\be
\label{cp}
\hat{\cal H} \, | \Psi_{\! \rm planck}\rangle \, = \, 
\sqrt{-g}(\kappa R - {1\over 2} (\nabla \phi)^2) \;
|\Psi_{\! \rm planck}\rangle.
\ee
Here we used that the one-loop term, proportional to $\hbar\, ({4\over
3} W \! - {1\over 2} \partial_\phi^2 W)$, is cancelled by the
corresponding fermionic contribution, which is true provided the
fermionic part of the wavefunction is chosen such that $|\Psi_{\rm
planck}\rangle$ indeed represents a supersymmetric ground state.
Combined with the $\hat{\cal H}=0$ condition (\ref{cst}), eqn
(\ref{cp}) implies 
\be
\label{vijf}
\Bigl\langle \, 
\kappa R - {{1\over 2}} (\nabla \phi)^2 \, \Bigr \rangle 
\, = \, {1\over Z} \langle \Psi_{\! \rm matter} | \, 
\widehat{\cal U} \, \hat{\cal H}\, |\Psi_{\! \rm 
planck}\rangle =\, 0.
\ee
This equation generalizes the classical result of section 3. 
As emphasized before, possible corrections to this identity can only
arise from matter and/or supersymmetry breaking terms 
at the  direct location of the Planck brane.

\newsubsection{Supersymmetry breaking}

Suppose we insist on maintaining the Planck brane as supersymmetric as needed to make the
cosmological constant as small as observed today.  The question then arises as to whether the
evolution into the 5-d bulk, that describes the holographic RG flow towards the IR, can still
lead to a realistic non-supersymmetric low energy field theory on the matter brane.  A possible
scenario by which this may happen is when the RG flow at some scale enters a strongly coupled
region (coupling $g$ of order 1 or larger) from both the 5-d and 4-d point of view.  The
dynamics in this region could then be sufficiently non-trivial to induce dynamical breaking of
4-d supersymmetry.

\smallskip

To gain some insight into how this may happen, suppose
we define the matter state by means of some non-supersymmetric 
low energy matter sector, so that 
\be
\label{nonsusy}
Q_\alpha |\Psi_{\rm matter} \rangle \, \neq \, 0.
\ee
The vacuum amplitude is then still supersymmetric,
since the Planck state
via (\ref{z}) projects out the supersymmetric part of the matter wave-function.
The violation of supersymmetry becomes visible, however, as soon as we consider
non-trivial expectation values. Generalizing our prescription of section 4, 
we can formally represent these via
\be
\label{mat}
\langle \Psi_{\rm matter} | \, {\cal O}_{phys} \, \widehat{\cal U} \, |\Psi_{\rm
planck} \rangle
\ee
where ${\cal O}_{phys}$ represents some physical observable inside the matter brane
region. Suppose we now try to derive a supersymmetry Ward identity. 
Because of (\ref{nonsusy}), it will be violated
\be
\langle \Psi_{\rm matter} | \, [ Q_\alpha ,\, 
{\cal O}_{phys} ]  \, \widehat{\cal U} \, |\Psi_{\rm
planck} \rangle \, \neq \, 0.
\ee
The 4-d low energy observer would thus conclude that supersymmetry is broken.  On the other
hand, since physical operators ${\cal O}_{phys}$ must commute with the generator $\hat{\cal
H}$,  the derivation of eqn (\ref{vijf}) as given in the previous section
still goes through (see also Appendix A).

\newsubsection{4-d Effective Field Theory}

The reasoning in the above sections is rather formal.  To make things somewhat more
concrete, let us from now on assume that the 4-d high energy gauge theory is at large $N$ and
strong 't Hooft coupling, so that the 5-d supergravity is in its classical regime.

\smallskip

In terms of pure 4-d language the situation is then summarized as follows. 
Consider the complete matter effective action, computed in
some {\it fixed} 4-d background $(g,\phi)$, obtained by integrating
out both the low energy energy matter, as well as the high energy matter
dual to the 5-d bulk supergravity. We will still call this
effective action $\Gamma_{\rm matter}$, since indeed it is just 
the same action (\ref{gamma}) evolved via the RG flow (\ref{cst}) 
to a larger scale factor for $g_{\mu\mu}$.
Note that $\Gamma_{\rm matter}$ now contains the full 
Einstein term of the 4-d gravity. The total 4-d action is 
obtained by just adding the Planck brane action (\ref{fine})
\be 
\label{gammat}
\Gamma_{\rm total}(g,\phi) = \Gamma_{\rm matter}(g,\phi) + 
\int \! d^4 x \, \sqrt{-g} \, W(\phi).
\ee
The claim is that, regardless of the details of the low energy matter 
theory, the classical equations of motion of this action
\ba
{1\over \sqrt{-g}}
{\delta \Gamma_{\rm matter} \over \delta g^{\mu\nu} } \is {1\over 2} 
W(\phi) \, g_{\mu\nu}
\nonumber \\[3.5mm]%\qquad \qquad \ 
{1\over \sqrt{-g}}
{\delta  \Gamma_{\rm matter} \over \delta \phi} \is \partial_\phi W(\phi)
\label{eom}
\ea
are such that the trace of the Einstein equation always takes the
form (\ref{vier}). This is possible since $\Gamma_{\rm matter}$ is 
not completely arbitrary, but due to the high energy
equivalence with the 5-d supergravity satisfies 
(\ref{cst})-(\ref{fff}). In the classical approximation, 
this identity reduces to the Hamilton-Jacobi relation, which 
looks identical to eqn. (\ref{drie}) with the replacement
\ba
\theta_{\mu\nu} \! - \theta g_{\mu\nu} \is  {1\over \sqrt{-g}}
{\delta \Gamma_{\rm matter} \over \delta g^{\mu\nu} }\nonumber
\\[3.5mm]
%\qquad \qquad \ 
\partial_n \phi \is {1\over \sqrt{-g}}
{\delta  \Gamma_{\rm matter} \over \delta \phi}.
\label{replace}
\ea
Via this same replacement, the classical equations of motion (\ref{eom})
just amount to the boundary conditions (\ref{twee}). Combining these two results 
gives (\ref{vier}), just as before.

\newsubsection{Stabilizing the Matter Brane}

Finally, let us address how to stabilize the matter brane.  This is an important
issue, since its location relative to the Planck brane represents an invariant
physical scale \cite{rs1}.  As expected, some subtleties will arise here.

\smallskip

A first subtlety is that the 5-d physics in the matter brane region is likely to
be singular.  Thus far, however, we did not need to worry about this issue,
because we chose to describe the matter brane region by means of the dual low
energy field theory.  No special assumptions about the properties of the
singularity were needed, except that at some distance away from it we can use
the geometric supergravity language to derive the constraint (\ref{cst}).  It is
not important for our argument, for example, whether there are discrete or
continuously many solutions to this constraint.

\smallskip

Regardless of this issue, it is clear that the matter brane effective action
should be allowed to be as arbitrary as possible.  Nonetheless one would hope to
naturally stabilize its location by means of its interaction with the Planck
brane.  We will first describe two possible mechanisms, which however both will
fall short in that they either remain unstable or need unnatural fine-tuning.
We will then describe a third scenario that will resolve these problems.

\smallskip

\smallskip

\bigskip

\noindent
{\it Attempt 1: Goldberger-Wise mechanism \cite{gw}}

\smallskip

Suppose that, in spite of the fact that it describes a singular space-time 
region, we would assume that the matter brane dynamics is well approximated 
by that of some classical brane with some arbitrary tension $\Lambda(\phi)$. 
We can then look for a classically stable location for matter brane as 
follows \cite{st}. A static bulk solution, when matched onto the supersymmetric boundary conditions set
by the Planck brane, is described by scalar fields $\phi(r)$ and a metric 
\be 
\label{dee}
ds^2 =\,  a^2(r) \, \eta_{\mu\nu} dx^\mu dx^\nu +
dr^2, 
\ee 
satisfying the supersymmetric flow equations 
\be
\label{hrg1} 
{a'\over a}
= -{1\over 6} W(\phi) \qquad \qquad \phi' = \partial_\phi W.  
\ee 
From this one finds that the matching relations at the matter brane, that are
required for there to be a flat solution, are that (cf. eqn (\ref{eom}))
\be
\label{mmm}
\partial_\phi \Lambda(\phi_c) = \partial_\phi W(\phi_c) \qquad \qquad \quad \
\Lambda(\phi_c) = W(\phi_c) 
\ee
for some critical value of $\phi_c$ for the scalar fields. The first
equation generically gives at most a discrete set of solutions
for $\phi_c$. The second relation, however, is then valid only 
if the value of $\Lambda$ at such a critical point $\phi_c$ is exactly
equal to that of the superpotential $W$. This amounts to an 
unnatural fine-tuning of the matter brane action. However, even
if we would choose $\Lambda(\phi)$ such that this condition is satisfied, 
the equations of motion (\ref{mmm}) still fail to stabilize the relative 
location of the Planck and matter brane, as they do not pick out
one particular preferred relative ratio for the scale factors $a$
at the two branes.

\smallskip

Because of both these problems, we will instead choose a different route.
%which will be more in accordance with the discussion as given in the previous sections.

\bigskip

\smallskip
\smallskip

\noindent
{\it Attempt 2: Hamilton-Jacobi action}

\smallskip

In accordance with our definition of the matter brane as the region behind the 
dashed line in fig 1, we will now reinstate the condition that its 
effective action satisfies (\ref{cst}). Let us further assume that for
slowly varying fields it takes the general form
\be
\label{low}
\Gamma_{\rm matter} = \int \! \sqrt{-g} \, (\Lambda(\phi)\, +\, 
\Phi(\phi) R\, +\, \ldots)
\ee
Using the classical supergravity approximation, (\ref{cst}) then 
reduces to the Hamilton-Jacobi relation \cite{rgt}
\be
\label{hhj}
{1\over 2} (\partial_\phi \Lambda)^2  \! -{1\over 3} \Lambda^2
\; + \; \Bigl(
\partial_\phi \Lambda \, \partial_\phi \Phi \! 
- {1\over 3} \Lambda \Phi)\, R  \, + \, \ldots  
= \,
V \, + \, \kappa \, R\,  + \, \ldots
\ee
with $V$ and $\kappa$ the 5-d potential and Newton constant. Apart
from this constraint, the functions $\Lambda(\phi)$ and $\Phi(\phi)$
will be allowed to be arbitrary.  

\smallskip
In principle, one could read (\ref{hhj}) as a condition on classical
field configurations $(g_c,\phi_c)$, that determines the matter brane curvature $R$
for given tension $\Lambda(\phi_c)$, etc. In our set-up, 
however, (\ref{low}) defines a true Hamilton-Jacobi action,
for which (\ref{hhj}) in fact amounts to a {\it functional identity} valid
for {\it all} field configurations $(g, \phi)$. So in particular the relation
\be
\label{up}
{1\over 2} (\partial_\phi \Lambda)^2  \! -{1\over 3} \Lambda^2 = V
\ee
represents a relation between $\Lambda(\phi)$ and $V(\phi)$ that holds for all 
values of $\phi$.

\smallskip
 
Next let us again look for a
consistent flat matter brane solution in the presence of the Planck
brane. Following the same derivation as before, we again arrive at the
condition (\ref{mmm}) for $\Lambda(\phi)$. 
Now the situation is somewhat different: if we
can find a solution to the first relation in (\ref{mmm}) then via
(\ref{up}) we automatically satisfy the second relation. At first this
seems like good news, but in fact the situation is now worse than
before: for generic $V(\phi)$ the only solution to the H-J relation
(\ref{up}) that will ever be tangent to $W(\phi)$ is $W(\phi)$
itself. Hence if we require that (\ref{mmm}) is satisfied, we
must have that $\Lambda(\phi) = W(\phi)$. But this is a hyper fine-tuned 
situation, since it says that the matter brane is just as
supersymmetric as the Planck brane. This is not what we want.

\smallskip

\smallskip
\smallskip

\bigskip

\noindent
{\it Attempt 3: Holographic effective action}

\smallskip

The problem we just encountered is in essence Weinberg's
counter-argument against the use of various adjustment
mechanisms for cancelling the cosmological constant \cite{weinberg}.
We now propose a mechanism that will evade this counter-argument.  
See also \cite{eh}.

\smallskip

There are many indications that the matter brane action $\Gamma_{\rm
matter}(\phi,g)$, at finite values for the scale factor of $g_{\mu\nu}$, 
in fact corresponds to a Wilsonian effective action of 
the dual 4-d field theory, defined with a finite UV cut-off $\epsilon$. 
Therefore, a more accurate formula for the matter effective action is 
\be
\label{loww}
\Gamma_{\rm matter} = \int \! \sqrt{-g} \, (\Lambda(\phi,\epsilon)\, + \,
\Phi(\phi,\epsilon) R \, + \, \ldots).
\ee
This cut-off dependence of $\Gamma_{\rm matter}$ will turn out to be crucial. 
The holographic dictionary furthermore relates variations in $\epsilon$ 
to variations in the scale factor of $g_{\mu\nu}$ via
\be
\label{deep}
\epsilon\, {\partial\  \over \partial \epsilon} \, = \, 2 \int \! 
g^{\mu\nu} {\delta\  \over \delta g^{\mu\nu}} \, = - \, a
{\partial \over \partial a}
\ee
with $a$ the overall warp factor in (\ref{dee}).  We can use this relation to
replace the short-distance cut-off $\epsilon$ in (\ref{loww}) by $a$.
The effective action (\ref{loww}) thus acquires a new and unexpected 
dependence on $g_{\mu\nu}$.

\smallskip

The holographic relation (\ref{deep}) between RG transformations
and constant Weyl rescalings, and the resulting extra dependence of $\Gamma_{\rm matter}$
on $g_{\mu\nu}$, has quite non-trivial implications. In particular, 
it implies that the Hamilton-Jacobi relation (\ref{up}) and matching relations 
(\ref{mmm}) now take the more general form:
\be
\label{up2}
{1\over 2} (\partial_\phi \Lambda)^2  -{1\over 48} \Bigl(4 \Lambda
-a {\partial\Lambda\over \partial a} \Bigr)^2\,
= \, V ,
\ee
\be
\label{mmm2}
\partial_\phi \Lambda(\phi_c,a) 
\, = \, \partial_\phi W(\phi_c), 
\ee
\be
\Bigl ( 4 - a {\partial \ \over \partial a} 
\Bigr)\Lambda(\phi_c,a) \, = \, 4 W(\phi_c). 
\label{mmm3}
\ee

\smallskip

\smallskip

\noindent
In fact, now we are in business: it is easy to show that these equations in
general {\it do} have solutions. The reason is that, via the non-trivial 
dependence on $a$, we now have an extra
parameter that we can adjust. 

\smallskip

Things are however still less trivial than
one might expect: the above equations have an RG invariance
and generically allow for a whole {\it critical line} of
solutions $\phi_c(a)$, generated by the holographic RG-flow 
(see eqn (\ref{hrg1})):
\be
a{\partial \over \partial a} \phi_c(a) = \beta_\phi(\phi_c(a)) 
\ee
\be
\label{beta}
\beta_\phi = {6 \partial_\phi W\over W}.
\ee
Eqn (\ref{mmm2}) does therefore not pick out one
particular preferred value for the scale $a$ at which the matter
brane is localized: $a$ is still left free. 
This is in accordance with the definition of the matter brane 
as the region behind a relatively arbitrary
line of separation with the bulk region. 

\smallskip

Nonetheless, we have achieved a stabilization of the matter brane
location.  For a given matter brane
effective action (\ref{loww}), eqn (\ref{mmm2}) selects out a
particular RG trajectory. This trajectory will generically become
unstable around a particular scale $a_{crit}$, where one or more of
the scalar fields starts to run off a steep slope and induce a large
5-d curvature. This critical scale $a_{crit}$ is the natural scale of
the matter brane. and also the scale where non-supersymmetric
physics can start to take place.

\newsubsection{An Explicit Example}

To illustrate the above stabilizing mechanism, we
now describe a specific example. 
Consider an effective potential of the matter brane of the form
\be
\label{lambd}
\Lambda(\phi,a) = - W(\phi) + \omega(\phi,a)
\ee
where $\omega(\phi,a)$ can be treated as a small perturbation. Eqns
(\ref{up2})-(\ref{mmm3}) then reduce to 
\be
\label{rgl}
\Bigl(a{\partial \over \partial a} - \beta_\phi\partial_\phi - 4\Bigr)
\, \omega(\phi,a ) \, = \, 0
\ee
\be
\partial_\phi \omega\, (\phi_c, a) \,=\, 0
\ee
\be 
\Bigl(a {\partial \over \partial a} -4\Bigr)\,  \omega\, (\phi_c, a) = 0,
\ee
with $\beta_\phi$ as defined in (\ref{beta}). The first equation
states that $\omega$ is invariant under the holographic RG flow
generated by $W$. The last two equations further show that $\omega$ is
the potential that needs to be minimized to obtain the classical
vacuum values of the fields. We can thus identify $\omega(\phi,a)$ 
with the total effective potential.

%\figuurb{plot}{5cm}{{\bf Fig 2.} \ {\it Graph of the effective potential $\omega(\phi,a)$ given in eqn (\ref{omeg}). 
%\\[-14mm]}}

\smallskip

Now as a simple example, let us consider the case with just one scalar field $\phi$
-- the dilaton -- and assume that 
\be
W(\phi) = e^{\textstyle {1\over \sqrt{3}} \, {\phi}}.%/\sqrt{3}}
\ee
%with $\alpha$ some constant. The corresponding holographic beta-function 
is 
\be
\beta_\phi = 2\sqrt{3}.
\ee
Now suppose that $\omega(\phi,a)$ takes the form\footnote{
This potential has been deliberately chosen to take
the typical form of a one-loop correction, obtained by integrating out
a 4-d matter field, in this particular case with a mass proportional to 
$e^{{1\over 2\sqrt{3}}\phi}$. The extra $a$-dependence of the potential
can be derived for example by computing the one-loop determinant
via dimensional regularization, in the 4-d induced background 
metric $g_{\mu\nu} = a^2 \eta_{\mu\nu}$ on the matter brane.} 
\be
\label{omeg}
a^4 \, \omega(\phi,a) \, = 
%\, - {1\over 2} \, \mu \, a^{2}\;
%e^{\textstyle{{1\over \sqrt{3}} \, \phi}}%/\sqrt{3}} 
{1\over 4} \, \lambda \, a^{4} \;  e^{\, \textstyle{
{2\over \sqrt{3}}\, \phi}} \, \Bigl(\; \log\Bigl(
 \mu\, a^{2}\; e^{\textstyle{{1\over \sqrt{3}}\, \phi}}\, \Bigr) \, - \,
{1\over 2}\, 
\Bigr) 
%/\sqrt{3}} 
\ee
with $\lambda$ and $\mu$ some free parameters. It is easily verified that 
this expression, for arbitrary $\lambda$ and $\mu$,  solves the linearized 
RG flow equation (\ref{rgl}).

\smallskip

The minimum of the effective potential lies at
\be
%\qquad \qquad 
%\qquad a^{2} \; e^{\textstyle {1\over \sqrt{3}} \, \phi} \, \is \, 0 \qquad \qquad 
%\qquad \qquad \ \ \; T> T_C \nonumber \\[4mm]
%\qquad \qquad \qquad 
\mu a^{2} \; e^{\textstyle{1\over \sqrt{3}} \, \phi_c} 
%+ \textstyle{1\over2}
\, = \, % \is \,  
1 %\over \lambda} .\, %(T_C-T \, )
%\qquad \qquad \ T< T_C \nonumber
\ee
This minimum describes an RG trajectory 
\be
\label{tra}
\phi_c(a) = -\sqrt{3}  \log ( {\mu   a^2} )
.
\ee
Via (\ref{hrg1}) this corresponds to a classical bulk geometry (\ref{dee}), with
warp factor
\be
\label{wcl}
a(r) = \sqrt{1 - {r \over 3 \mu}} \qquad \qquad 0 \leq r \leq {3
\mu}. 
\ee
Here we normalized $a$ such that the Planck brane is located at $a=1$.
We see that the $r$-direction describes a compact interval: the geometry closes off by means
of a naked singularity, located at the critical distance 
$r_{crit} = {3\mu}$ from the Planck brane. 

\smallskip

The matter brane is the strongly curved neighborhood of the naked
singularity. Its location is therefore indeed dynamically stabilized by the
presence of the Planck brane. Furthermore, by virtue of our 
presumption that the Planck brane remains supersymmetric, both branes are flat.
This result holds for arbitrary values of the free parameters $\mu$ and $\lambda$.
Note that the effective 4-d cosmological constant of the matter brane is indeed
{\it not} equal to the minimum value of the effective potential %$\omega(\phi,a)$
\be
\label{mini}
a^4 \omega(\phi_c,a) = -{\lambda \over 8\mu^2}.
\ee
Rather, it is equal to its variation with respect to the overal scale $a$ 
\be
\Lambda_{\rm eff} \, = \, a {d\over d a} \Bigl( a^4 \omega(\phi_c,a)\Bigl) \, 
= \, 0.
\ee 
This last equality of course follows by inspection from (\ref{mini}), but more fundamentally,
arises as a direct consequence of the (linearized) RG constraint (\ref{rgl}).

\smallskip

The RG trajectory (\ref{tra}) and geometry (\ref{wcl}) describe a
supersymmetric bulk solution.  This is the classical version of the result that in the 
vacuum amplitude (\ref{z}), the Planck state always projects out the supersymmetric component of the matter
state.  We expect, however, that for general non-supersymmetric observables, the non-supersymmetric
form (\ref{lambd})-(\ref{omeg}) of the matter brane effective action will become manifest.
Supersymmetry remains intact only for the BPS-type limit $\lambda \to 0$ 
and $\mu \to 0$ (in some fixed ratio).

\bigskip

\newsubsection{Discussion}

We have explored some physical consequences of a Randall-Sundrum scenario
\cite{rs1}, in which the Planck brane region is assumed to be supersymmetric.
Via the UV/IR mapping of the AdS/CFT duality, this condition appears equivalent
to that of 4-d high energy supersymmetry.  As we have shown, however, the same
supersymmetry also stabilizes the large scale 4-d geometry.  It appears
therefore that our set-up implies a stronger restriction on the 4-d effective
field theory than ordinary high energy supersymmetry.  From the higher
dimensional viewpoint this stronger restriction seems natural, however, as it
pertains to just one single region of the 5-d geometry.\footnote{In
\cite{witcos}, a somewhat similar suggestion was made for a scenario in which
our 4-d world might be stabilized via the embedding into an unobservable 5-d
supersymmetric world.}

\smallskip

Our set-up differs in a number of aspects from other recent proposals \cite{adks} \cite{kss}.
In these papers, the effective 4-d flatness appears as a consequence of a self-tuning
mechanism of a single 4-d brane-world, embedded inside a fine-tuned 5-d bulk.  This mechanism
however relies on rather strong assumptions about the physics of the IR singularity of the 5-d
geometry.  In our case, on the other hand, the 4-d stability is the consequence of a
fine-tuned Planck brane action.  This fine-tuning is justified by our assumption of
supersymmetry, and furthermore {\it local}, both from the 5-d and the 4-d RG perspective.

\smallskip
  
Given our observation that the 4-d flatness is proportional to the 
amount of supersymmetry breaking at the Planck brane, it will clearly be
important to look for possible mechanisms that may prevent or suppress the mediation of
supersymmetry breaking via the bulk.  It seems clear that holography (both 5-d and 4-d) will
be an important ingredient in this discussion. 

\smallskip

First, it will be important to know how best to describe the local degrees of freedom on the
Planck brane.  Via the string theory description \cite{hv} \cite{cpv}, much is known in
principle about its internal structure, but its effective dynamics still seems hard to
extract.  Let us nonetheless imagine computing some supersymmetry violating one loop
correction to the Plank brane physics, induced by a bulk graviton traveling back and forth to
the matter brane region.  A priori, one would not expect any large suppression from this
propagator, and it would thus appear that supersymmetry will indeed be violated on the Planck
brane at roughly the same length scale as that set by the matter brane location.

\smallskip

This conclusion, however, disregards the implications of 5-d holography, which
suggest that the local quantum fluctuations at the Planck brane -- in as far
present -- all have roughly Planckian frequencies and wave-lengths.  If this is
the case, the 5-d graviton modes emitted by these fluctuations must also have
Planckian wavelengths along the brane direction.  Such modes, however, decay
exponentially along the $r$-direction with a Planckian length.  Hence, if this
physical picture is correct, this would indeed give a mechanism by which the
bulk mediation of supersymmetry breaking gets exponentially suppressed via the
separation between the two branes.

\smallskip

Additional restrictions on the use of naive effective field theory may come from
4-d holography.  It is clear, for example, that a direct AdS/CFT interpretation
of the 5-d bulk region, as dual to a 4-d quantum field theory with Planckian
cut-off, still introduces far too many degrees of freedom.  It is tempting to
speculate that supersymmetry near the Planck region could be instrumental in
hiding this excess in degrees of freedom, thus protecting the 4-d
holographic bound.

\figuur{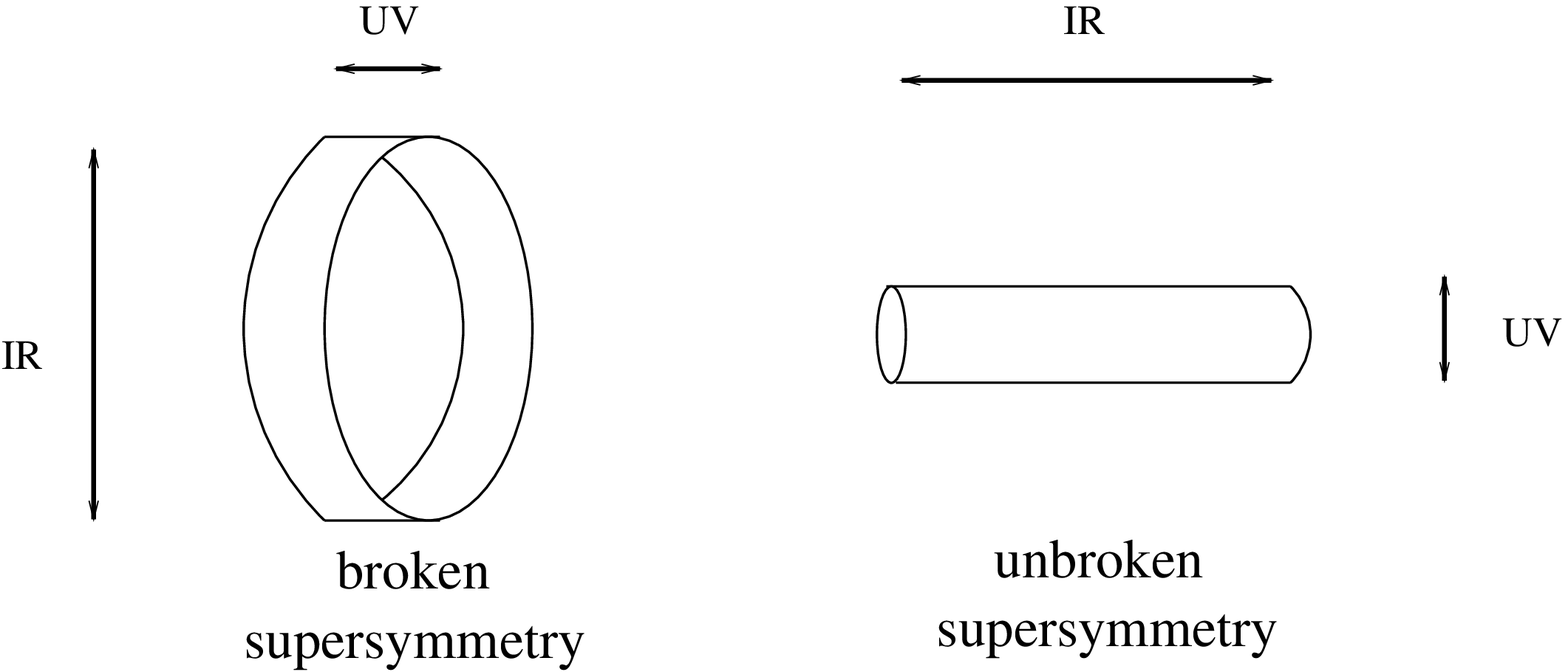}{11cm}{{\bf Fig 2. } \ {\it The geometric distinction between the matter and Planck regions
in fig 1, when translated in terms of the shape of string world-sheets, suggests that while 
large open string loops are non-supersymmetric, long closed string propagators can still be protected by 
supersymmetry.
\\[-5mm]}}

\vspace{-4mm}

Finally: the UV/IR connection explored here, just like all other ones pointed out
earlier in \cite{dkps} \cite{noncom} \cite{kv}, seems directly
connected with the fundamental equivalence between open string loops 
and closed string propagators.  Roughly, 
our assumption here has been that
long closed string propagators and short open string loops are protected by supersymmetry, 
even though short closed string propagators and long open string loops are not. 
It is reassuring that these two types of diagrams represent clearly separated 
regions in the parameter space of open string Feynman diagrams.

\bigskip

\noindent
{\sc Acknowledgements} \\
This work is supported by NSF-grant
98-02484.  I would like to thank Micha Berkooz, Lisa Randall,  Raman Sundrum
and Erik Verlinde for helpful discussions.

\bigskip

\renewcommand{\theequation}{A.\arabic{equation}}
\setcounter{equation}{0}

\figuurb{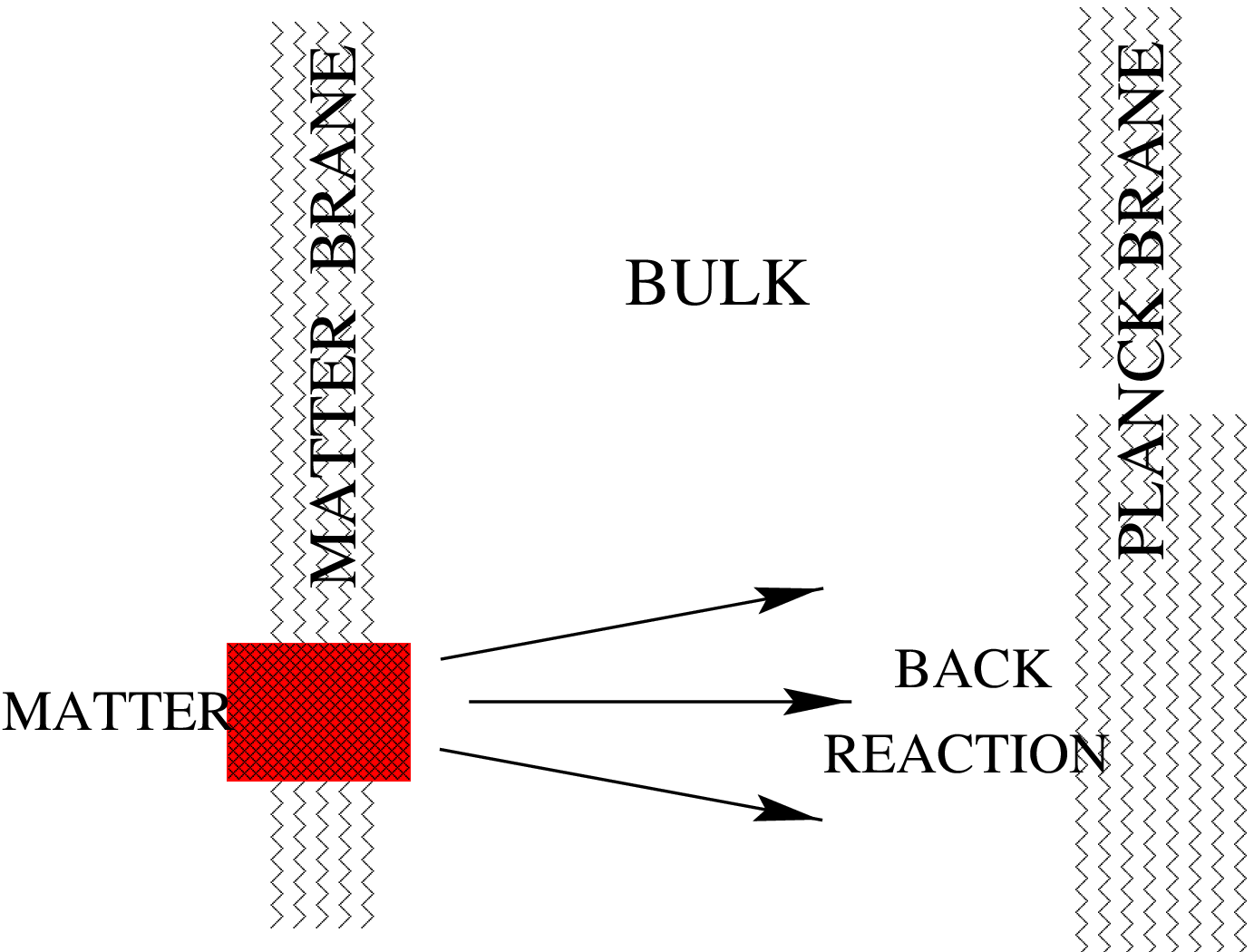}{6.8cm}{{\bf Fig 3. } \ {\it The backreaction on the geometry of a matter 
perturbation.\\[-14mm]}}

\vspace{-4mm}

\noindent
{\bf Appendix: A Small Matter Perturbation}

\smallskip

Here we consider the effect of a small matter perturbation 
on the geometry of the Planck brane. The perturbation 
induces a small extra contribution 
$\delta \, \Gamma(g,\phi)$ to the matter+bulk effective action.
The variation of this extra contribution with respect to the metric and
field $\phi$ represent the expectation values of the corresponding
dual operators
\be
{1\over \sqrt {-g}} {\delta (\delta\, \Gamma) \over \delta g^{\mu\nu}}
\, = \, \langle\, T_{\mu\nu} \rangle \qquad \qquad \
{1\over \sqrt {-g}} {\delta (\delta \, \Gamma) \over \delta \phi}
\, = \, \langle \, {\cal O}_\phi\,  \rangle
\ee
which represent the physical effect of the matter perturbation.
The RG flow relation (\ref{cst}) 
must still hold for the total effective action,
including the extra term. (This is true since we assume that all
matter is located on the matter brane.) 
This RG condition implies that the above
two expectation values cannot be independent. Working to linearized
order in $\langle\, T_{\mu\nu}\rangle$, and within the classical
supergravity approximation, one finds that 
\be
\label{reln}
\langle\, T \, \rangle\,  = \,  \beta_\phi \, 
\langle \, {\cal O}_\phi\, \rangle \; + \; {6\over W} \, 
\langle \, {\cal O}_\phi\,  \rangle \, \langle \, {\cal O}_\phi\, \rangle  
\ee
%\be\beta_\phi = - {6 \partial_\phi W\over W}  \qquad \qquad \gamma = {W\over 6}\ee    

Here $T$ is the trace of the stress energy tensor
and $\beta_\phi$ the holographic beta-function defined in (\ref{beta}).
Now if we consider the linearized equations of motion of the small metric 
variation $h_{\mu\nu} = \delta g_{\mu\nu}$ induced by the matter 
perturbation, then because of the relation (\ref{reln})
%between $\langle \, T\, \rangle $ and $\langle \, {\cal O}_\phi \, \rangle $, 
it is possible to combine the trace of the $h$ equation with the $\phi$ 
equation such that the source term $\langle \, T \, \rangle$ cancels. 
The resulting equation takes the form 
\be
\kappa\, 
\mbox{\large \raisebox{-1pt}{$\Box$}}  h  + {1\over 2} (\nabla\phi)^2 = 0.
\ee
The trace of the stress energy tensor of matter away from the Planck brane
thus always gets represented by means of variations in the $\phi$ field.

\renewcommand{\Large}{\large}

\bigskip

\noindent
\end{document}